# Highly efficient and tuneable spin-to-charge conversion through Rashba coupling at oxide interfaces

E. Lesne[1][♦], Y. Fu[2][♦], S. Oyarzun[2][♣], J.C. Rojas-Sanchez[1][♣], D.C. Vaz[1], H. Naganuma[1,3], G. Sicoli[2], J.-P. Attané[2], M. Jamet[2], E. Jacquet[1], J.-M. George[1], A. Barthélémy[1], H. Jaffrès[1], A. Fert[1], M. Bibes[1]* and L. Vila[2]

[1] Unité Mixte de Physique, CNRS, Thales, Univ. Paris-Sud, Université Paris-Saclay, 91767 Palaiseau, France

[2] INAC/SP2M, CEA-UJF, 38054 Grenoble, France

[3] Tohoku University, Department of Applied Physics, 6-6-05 Aoba, Aramaki, Aoba, Sendai 980-8579, Japan

**The spin-orbit interaction couples the electrons' motion to their spin. Accordingly, passing a current in a material with strong spin-orbit coupling generates a transverse spin current (spin Hall effect, SHE) and vice-versa (inverse spin Hall effect, ISHE)[1–3]. The emergence of SHE and ISHE as charge-to-spin interconversion mechanisms offers a variety of novel spintronics functionalities[4,5] and devices, some of which do not require any ferromagnetic material[6]. However, the interconversion efficiency of SHE and ISHE (spin Hall angle) is a bulk property that rarely exceeds ten percent, and does not take advantage of interfacial and low-dimensional effects otherwise ubiquitous in spintronics hetero- and mesostructures. Here, we make use of an interface-driven spin-orbit coupling mechanism – the Rashba effect[7] – in the oxide two-dimensional electron system (2DES) $LaAlO_3/SrTiO_3$ to achieve spin-to-charge conversion with unprecedented efficiency. Through spin-pumping, we inject a spin current from a NiFe film into the oxide 2DES and detect the resulting charge current, which can be strongly modulated by a gate voltage. We discuss the amplitude of the effect and its gate dependence on the basis of the electronic structure of the 2DES.**

[♦] These authors contributed equally to this work

* manuel.bibes@cnrs-thales.fr

[♣] Current affiliation : Institut Jean Lamour (UMR CNRS 7198), Université de Lorraine, 54500 Vandoeuvre-lès-Nancy, France; juan-carlos.rojas-sanchez@univ-lorraine.fr

[♠] Current affiliation : Departamento de Física, Universidad de Santiago de Chile (USACH), Chile

Perovskite oxide materials possess a broad range of functionalities, some of which can be very appealing for spintronics. This includes half-metallicity in mixed-valence manganites – that can be used to produce giant tunnel magnetoresistance[8] – or multiferroicity – through which magnetization direction can be electrically controlled at low power[9]. The recent years have seen the emergence of novel spintronics effects based on the generation and control of *pure* spin currents through spin-orbit effects in semi-conducting and metallic systems[1–3]. However, despite a renewal of interest for 4$d$ and 5$d$ transition metal perovksites[10], spin-orbit effects remained largely unexplored in oxide spintronics.

An emerging direction in oxide research aims at discovering novel electronic phases at interfaces between two oxide materials[11]. A well-known example is the LaAlO$_3$/SrTiO$_3$ system: while both LaAlO$_3$ (LAO) and SrTiO$_3$ (STO) are wide bandgap semiconductors, a high-mobility two-dimensional electron system (2DES) forms at their interface[12] if the LAO thickness is at least 4 unit-cells (uc). Interestingly, LAO/STO possesses several remarkable extra functionalities including a gate-tuneable Rashba effect[13,14], which makes it particularly appealing for spintronics.

The Rashba effect is a manifestation of the spin-orbit interaction (SOI) in solids, where spin degeneracy associated with the spatial inversion symmetry is lifted due to a symmetry-breaking electric field normal to an heterointerface[15]. In a Rashba 2DES, the flow of a charge current results in the creation of a non-zero spin accumulation[16,17] coming from uncompensated spin-textured Fermi surfaces. Recently, the converse effect so-called inverse Edelstein effect (IEE) – that is a spin-to-charge conversion through SOI – was discovered at Ag/Bi(111) interfaces[18]. Here, we report the observation of a very large, gate-tuneable IEE in NiFe/LAO//STO heterostructures, with a spin-to-charge conversion efficiency more than one order of magnitude larger than at the Ag/Bi(111) interface or in spin Hall materials.

NiFe/LAO//STO samples were grown in a setup combining pulsed laser deposition (PLD) and sputtering chambers with *in vacuo* sample transfer. Following our previous report on Co-capped LAO//STO samples[19], we exploit the possibility to reduce the critical LAO thickness required for 2DES formation below 4 uc through metal capping. Starting with TiO$_2$-terminated STO single crystal substrates, we first grow a 2 uc thick LAO film by PLD. As visible on the reflection high-energy electron diffraction (RHEED) images and intensity oscillations (Fig. 1a), the growth proceeds layer by layer. We then deposit a permalloy (Ni$_{81}$Fe$_{19}$) film (2.5 or 20 nm thick) by dc sputtering and cap it with aluminum (for transport experiments) or gold (for transmission electron microscopy, TEM). The sample surface is smooth, retaining the step-and-terrace topography of the STO and LAO//STO (cf. Fig.1b). TEM images (Fig. 1c) confirm the good crystallinity of the 2 uc LAO layer and indicate a nanocrystalline structure for the 2 nm

NiFe film and the Au cap. The interface between LAO and NiFe is smooth, with limited interaction between the two materials.

In Fig. 1d we compare the temperature dependence of the sheet resistance ($R_s$ vs T) for a typical NiFe(2.5 nm)/LAO(2 uc)//STO sample and a reference NiFe(2.5 nm)//STO single film. While the latter shows a weak resistance change between room and low temperatures, the former exhibits a clear resistance decrease. These transport data indicate the presence of an additional conductive channel in the NiFe/LAO(2 uc)//STO sample just as in Co/LAO(1-2 uc)//STO samples[19]. Assuming parallel conduction between the NiFe metal layer and this extra channel, we extract the resistivity vs temperature of the 2DES present in STO that is reminiscent of conventional LAO/STO 2DES behavior (cf. the *R* vs *T* data of a standard uncapped LAO(5 uc)//STO sample in the inset). From the analysis of Hall and magnetoresistance data in a parallel resistance model, we extract a total carrier density in the 2DES of $n_{total} \approx 2.6 \cdot 10^{13}$ cm$^{-2}$ ($n_{total}=n_1+n_2$, $n_1=1.54 \cdot 10^{13}$ cm$^{-2}$ and $n_2=1.07 \cdot 10^{13}$ cm$^{-2}$, with mobilities $\mu_1=3300$ cm²/Vs and $\mu_2=160$ cm²/Vs). We conclude that a 2DES is formed in STO in our NiFe(2.5 nm)/LAO(2 uc)//STO heterostructures, hence enabling the prospect of spin-pumping[20] from NiFe into the 2DES.

Prior to low temperature spin-pumping experiments, we characterized the dynamic magnetic response of our samples using ferromagnetic resonance (FMR) at room temperature. Fig. 2a displays the field derivative of the imaginary part of the dynamic magnetic susceptibility $\chi''$ for different frequencies *f* of the microwave excitation, as a function of the external magnetic field $H_{dc}$. For a NiFe(20 nm)/LAO(2 uc)//STO sample and a NiFe(20 nm)//LAO reference, we obtain the frequency dependence of the resonance field $H_{res}$ and the peak-to-peak linewidth $\Delta H_{pp}$ (cf. Fig. 2b and 2c, respectively). The dispersion relation of $H_{res}$ with $\omega = 2\pi f$ (Fig. 2b) follows the trend given by Kittel's relation :

$$\left(\frac{\omega}{\gamma}\right)^2 = (H_{res})(\mu_0 M_{eff} + H_{res}) \qquad (1)$$

with $\mu_0$ the vacuum permeability, $M_{eff}$ the effective saturation magnetization and $\gamma = g^* \mu_B/\hbar$, where $g^*$ is the effective electron *g*-factor (taken to be 2.11 for NiFe), $\hbar$ is the reduced Planck constant and $\mu_B$ the Bohr magneton. From Fig. 2c we also determine the effective Gilbert-like damping parameter $\alpha_{NiFe/2DES}$ according to

$$\Delta H_{pp}(f) = \Delta H_{inh} + \frac{2}{\sqrt{3}}\left(\frac{2\pi f}{g^* \mu_B}\right) \hbar \alpha_{NiFe/2DES} \qquad (2)$$

where $\Delta H_{inh}$ is the frequency-independent linewidth contribution (arising from inhomogeneities in the ferromagnet − FM). The effective damping for the NiFe(20 nm)/LAO(2 uc)//STO sample is enhanced compared to the reference ($\alpha_{NiFe/2DES}$=7.77x10$^{-3}$ vs $\alpha_{NiFe}$=6.45x10$^{-3}$), which is consistent with spin absorption/dissipation in the 2DES[20]. Nevertheless we do not observe spin-to-charge current conversion at room temperature, probably due to the short relaxation time.

Next, we make use of precessional excitation of the NiFe magnetization at T=7 K to inject a spin current into the LAO/STO 2DES (spin-pumping[20]), and we measure the voltage generated into the 2DES (spin-to-charge conversion) in the geometry sketched in Fig. 3a. In spin-pumping experiments, a prerequisite to the production of a sizeable charge current is the transmission of the out-of-equilibrium spin accumulation generated on the FM metal side to the non-magnetic (NM) layer. Whereas in all-metal systems, the spin accumulation easily diffuses from the FM to the NM this diffusion is impeded if a thin insulating layer is inserted at the FM/NM interface[21]. If the NM is a 2DES as here, the strongly localized character of the wavefunctions may however allow an exchange coupling between the FM and the NM and enable a substantial non-equilibrium spin accumulation to appear in the 2DES. Alternatively, spin-accumulation may also be produced in the 2DES through spin conduction by hopping processes via localized states with long spin-lifetime in the LAO (see Supplementary Information for details).

Fig. 3b displays the FMR response for positive and negative $H_{dc}$ at 7 K and a back-gate voltage of +125 V and Fig. 3c shows the detected raw voltage (normalized by the microwave power) and its decomposition into symmetric ($V_{sym}$) and antisymmetric ($V_{asym}$) components, for positive and negative $H_{dc}$. The symmetric component largely dominates the signal and is almost perfectly reversed upon reversing $H_{dc}$, as expected from spin-to-charge conversion effects with which $V_{sym} \propto j_s \times \sigma$ ($j_s$ being the spin current and $\sigma$ the spin direction that changes sign with the direction of $H_{dc}$). As in ISHE experiments, magnetotransport effects such as anisotropic magnetoresistance (AMR) or planar Hall effect (PHE) are likely responsible for the antisymmetric signal, and can also contribute to the symmetric part[22,23]. For comparison we show similar data for a reference NiFe(20 nm)//LAO sample (inset of Fig. 3c): the detected voltage is almost perfectly antisymmetric, evidencing the absence of spin-to-charge conversion in this sample and ruling out any contribution from the NiFe//LAO interface to the data measured in NiFe/LAO//STO. We thus conclude that spin-pumping and spin-to-charge conversion effects in the 2DES are responsible for the signal reported in the main panel of Fig. 3c. The interfacial charge current density corresponding to this value of the symmetric component of the measured voltage is simply given by $j_C^{2D} = V_{sym}/(wRh_{rf}^2)$=53.3 mA/m.G² with R=35.25 Ω the resistance of the NiFe/LAO//STO heterostructure and $w$=0.4 mm the sample width. Remarkably this value of $j_C^{2D}$ is orders of magnitude larger than in previous reports.

The most likely mechanism giving rise to spin-to-charge conversion into a 2DES with a sizeable Rashba effect such as the LAO/STO system is the so-called inverse Edelstein effect (IEE). In this picture, the spin current pumped in the interfacial Rashba 2DES selectively relaxes into the two spin-orbit split Fermi surfaces (cf. Fig. 3d). In the IEE framework, this leads to a shift of the Fermi contours in momentum space (cf. Fig. 3e), thereby producing a lateral interfacial charge current density[18,24]. A quantitative analysis of the IEE requires estimating the pumped spin current density from the relation:

$$j_S = \frac{g^{\uparrow\downarrow}\gamma^2\hbar(\mu_0 h_{rf})^2}{8\pi\alpha_{NiFe/2DES}^2}\left[\frac{\mu_0 M_{eff}\gamma + \sqrt{(\mu_0 M_{eff}\gamma)^2 + 4\omega^2}}{(\mu_0 M_{eff}\gamma)^2 + 4\omega^2}\right]\left(\frac{2e}{\hbar}\right) \quad (3)$$

with $\gamma$ the gyromagnetic ratio in NiFe, $h_{rf}$ the microwave field amplitude. $g^{\uparrow\downarrow}$ is the real part of the spin-mixing conductance given by

$$g^{\uparrow\downarrow} = \frac{4\pi M_{eff} t_F}{g^*\mu_B}(\alpha_{NiFe/2DES} - \alpha_{NiFe}) \quad (4)$$

with $t_F$=20 nm the thickness of the NiFe film.

From the out-of-plane angular dependence of the FMR signal at 7 K, we can extract $\alpha_{NiFe/2DES}$ at this temperature (see Supplementary Information), which yields $\alpha_{NiFe/2DES}$=7.7·10$^{-3}$, comparable to the room temperature value. From Eqs. (3) and (4) we obtain $g^{\uparrow\downarrow}$=13.3 nm$^{-2}$, in the range of reported values for interfacial systems[18,25], and $j_s$=8.4 10$^6$ A.m$^{-2}$.G$^{-2}$. Combining this value with the measured two-dimensional charge current yields the spin-to-charge current conversion efficiency parameterized by the inverse Edelstein length $\lambda_{IEE} = j_C^{2D}/j_S$. For the data shown in Fig. 3b, we find $\lambda_{IEE}$=6.4 nm.

This value of $\lambda_{IEE}$ in our NiFe/LAO/STO samples is one order of magnitude larger than the values of 0.1-0.4 nm found at Bi/Ag interfaces[18] and other Rashba interfaces[26], and is also larger than $\lambda_{IEE}$=2.1 nm found at the surface of the topological insulator $\alpha$-Sn (Ref. [27]). This spin-to-charge conversion efficiency also compares favourably with what is measured in heavy metals. In these systems, the interconverted spin and charge currents are both three-dimensional and the figure of merit is the spin Hall angle $\theta_{SHE}$, which amounts to 0.056 in Pt, 0.12 in Ta and 0.37 in W (Ref. [3]). For comparison purposes, one can convert the value of $\theta_{SHE}$ into $\lambda_{IEE}$ through $\lambda_{IEE}=\theta_{SHE}l_{sf}$ ($l_{sf}$ is the spin diffusion length) which yields 0.2 nm for Pt, 0.3 nm for Ta and 0.43 nm for W, well below the $\lambda_{IEE}$ =6.4 nm in our LAO/STO system.

At a Rashba interface, in the simplified approximation of circular spin contours, $\lambda_{IEE}$ can be expressed as a function of the momentum relaxation time $\tau$ and the Rashba coefficient $\alpha_R$ [24]

$$\alpha_R = \frac{\hbar \lambda_{IEE}}{\tau} \quad (5)$$

Assuming a mobility $\mu$=4000-5000 cm²/Vs (cf. [28]) and $m^* \approx 2m_0$ ($m_0$=9.1·10$^{31}$ kg) puts $\alpha_R$ on the order of 1·10$^{-12}$ eVm. This is in the range of calculated values[29] and compatible with $\alpha_R$=2.6-3.5 10$^{-12}$ eVm extracted from weak antilocalization (WAL) magnetoresistance data[13,30] (we note however that WAL treats spin-orbit coupling as a perturbation instead of considering a spin-momentum locking as is the case in the IEE).

The large dielectric constant of the STO substrate and the relatively low carrier density of the LAO/STO 2DES makes it possible to use a back-gate voltage $V_g$ to modulate the 2DES carrier density (by 0.5-1·10$^{11}$ cm$^{-2}$/V, Ref. [31,32]) and electronic properties. As we show in Fig. 4a and 4b, back-gating has a dramatic influence on the spin-to-charge conversion efficiency. The detected charge current evolves from a moderate positive value at negative $V_g$ (charge depletion regime), to a small positive value at $V_g$=0 and then becomes negative for positive gate voltage, showing a maximum amplitude around +125 V. This is summarized by the gate dependence of $\lambda_{IEE}$ displayed in Fig. 4d, from which a crossover between positive to negative spin-to-charge conversion is clearly visible near $V_g$=0.

The gate dependence of the IEE signal in our samples is highly non-trivial and likely related to the multiband nature of the 2DES electronic structure, and the possibility to tune the 2DES across Lifshitz points with the gate[31]. As pointed out in several studies[31,33], at low carrier density the electrons occupy a single low-lying band with $d_{xy}$ character, whereas raising the Fermi level $E_F$ through the application of a gate voltage promotes the population of $d_{xz,yz}$ bands. The amplitude of $\alpha_R$ and the associated spin textures are strongly dependent on the relative energy and the orbital symmetry of these bands[29,31,34]. First-principles calculations indicate that for the low-lying $d_{xy}$ band $\alpha_R$ is weak and has a negative sign while for the $d_{xz,yz}$ bands $\alpha_R$ is positive and increases from the band bottom to the avoided crossing point with $d_{xy}$ (Ref. [29]). Accordingly, the influence of spin-orbit effects on charge and spin transport strongly depends on the position of the Fermi level, i.e. on the gate voltage. With a carrier density of ~2.6·10$^{13}$ cm$^{-2}$ at $V_g$=0, the Fermi levels in our sample sits just above the bottom of the lowest $d_{xz,yz}$ band (corresponding to 1.7-1.9·10$^{13}$ cm$^{-2}$, Ref. [31,32]) and both $d_{xy}$ and $d_{xz,yz}$ bands are populated. Because $\alpha_R$ has an opposite sign for the two populated bands[29], spin-to-charge conversion effects tend to compensate each other, consistent with the low value of $j_C^{2D}$ we measure at $V_g$=0.

At negative gate voltages, $E_F$ decreases and only the $d_{xy}$ band with weak $\alpha_R$ is populated, which is in line with our observation of an increase of $j_C^{2D}$ when increasing $V_g$ from 0 towards negative voltages, levelling off at a moderate value, consistent with the weak energy dependence of the Rashba splitting expected for this band. At positive gate voltages, $E_F$ increases and eventually reaches the avoided crossing between $d_{xy}$ and $d_{xz,yz}$ where $\alpha_R$ is the largest[31,33,35], and has a sign opposite to that of the $d_{xy}$ band. This maximum of the Rashba splitting was reported to occur at carrier densities n≈3.3·10$^{13}$ cm$^{-2}$ (Ref. [33]), which is accessible with back-gate voltages on the order of +100 V, consistent with the position of the maximum at +125 V in Fig. 4d. Beyond that point, the Rashba splitting decreases[31,33]. Although ideally, spin- and angle-resolved photoemission experiments would have to be performed to ascertain the band structure of the 2DES in our NiFe/LAO//STO samples, the above scenario explains our observation of an increased $j_C^{2D}$ (in absolute value) at positive gate voltage, and its maximum.

In summary, we have measured a very large spin-to-charge conversion efficiency in the two-dimensional electron system present at the interface between LaAlO$_3$ and SrTiO$_3$. Conversion occurs through the inverse Edelstein effect arising from the Rashba coupling present at the interface. Upon application of a gate voltage, the amplitude of the converted current can be modulated over one order or magnitude, and even changes sign. This can be interpreted in terms of a crossover between the occupancy of one to several bands with different orbital characters and different spin-orbit textures. Our results suggest that oxide interfaces have a strong potential for spintronics[36], both for the generation or detection of spin-currents through direct[16] or inverse Edelstein[18,24] effects, and for their electrical modulation, *à la* Datta & Das[37].

**ACKNOWLEDGMENTS**

Research at CNRS/Thales received support from the ERC Consolidator Grant #615759 "MINT" and the region Île-de-France DIM "Oxymore" (project NEIMO). Support from the ANR SOspin project is also acknowledged. HN was partly supported by the Leading Young Researcher Overseas Visit Program, JSPS Grant-in-Aid for Scientific Research (B) (#15H03548). Authors are grateful to Y. Kodama (Tohoku University, Japan) for TEM observation and to N. Reyren for his help at the early stage of the project.

## METHODS

**Sample growth.** The LaAlO$_3$ (LAO) films were grown by PLD on 5 mm × 5 mm TiO$_2$-terminated (001)-oriented SrTiO$_3$ (STO) substrates (from Crystec GmbH). A single-crystal LAO target was ablated by a KrF (248 nm) excimer laser at a repetition rate of 1 Hz and with a fluence of ∼ 1 J/cm$^{-2}$. The LAO deposition was performed in an oxygen partial pressure of 2.0·10$^{-4}$ mbar and at a substrate temperature of 730°C. The substrate-to-target distance was 63 mm. After LAO growth, the samples were then annealed for 30 min in about 400 mbar of oxygen at 500°C. Finally, the LAO/STO heterostructures were cooled at 25°C/min and kept in the same oxygen pressure for ∼ 30 to 60 min. PLD growth was followed by *in situ* deposition of a metallic Ni$_{81}$Fe$_{19}$ layer by dc magnetron sputtering at room temperature in a pure Ar atmosphere, capped by Au or Al. For the NiFe//LAO reference sample, the NiFe layer was deposited on a (001)-oriented LAO substrate and capped with Al.

**Transmission electron microscopy.** A specimen for TEM observation was prepared using a combination of mechanical polishing and ion milling. The energy for Ar ion milling was gradually decreased from 4 to 1 keV to minimize ion beam induced damage. The same strategy was applied for focused ion beam energy from 30 to 4 keV. TEM analysis was performed using TEMs of JEOL operated at 300 keV.

**dc transport.** The longitudinal and transverse dc transport properties of the samples were measured in a Quantum Design Dynacool system as a function of temperature and magnetic field (up to ±90 kOe). The carrier density and mobility in the 2DEG were calculated from the measured Hall and magnetoresistance data of a NiFe/LAO//STO sample and a NiFe//STO reference, following Ref.[38] and using one and two carrier types for the metal and the 2DES, respectively.

**FMR and voltage measurements with applied back gate.** Frequency dependent ferromagnetic resonance (Fig. 2) was performed at room temperature by using broadband technique involving a co-planar waveguide. During each measurement, the microwave frequency was fixed while the external magnetic field was swept. The intensity of each spectra in Fig. 2a is proportional to the field derivative of the imaginary part of the dynamic magnetic susceptibility ($d\chi''/dH$) obtained by using a lock-in technique and field modulation. By fitting the frequency dependences of the resonance field and linewidth, the effective magnetization $M_{eff}$ and the damping parameter were obtained, which excludes linewidth broadening due to frequency-independent inhomogeneities. For low temperature spin-pumping experiments, FMR and dc transverse voltage measurements were performed simultaneously on 2.4 mm × 0.4 mm slabs. To realize the electrical connections (schemed in Fig. 3a), a sample holder made of a printed circuit board (PCB) circuit was used. The two contacts at the metallic top layer were connected to the PCB sample holder via wire bonding, and the back gate was connected by gluing the substrate on one contact of the PCB sample holder using conductive silver paint. To perform the measurements, the sample was placed at the center of a cylindrical X-band cavity ($f \approx 9.6$ GHz, TE$_{011}$ mode). The system was cooled down from room temperature to 7 K using a Helium-flow cryostat. At 7 K, the sample was initialized by sweeping the back-gate voltage from +200 V to -200 V then back to +200 V, at which the first measurement was performed. The spin-pumping measurements were performed at fixed $V_g$ and $V_g$ was changed monotonously from +200 V to -200 V. The linear dependence of the measured dc voltage amplitude on the applied power of the microwave excitation has been verified up to around 3 mW at 7 K. In our measurements the microwave power was 1.99 mW. The quality factor Q of the cavity was measured to calculate the radio-frequency magnetic field amplitude $h_{rf}$ that was used to calculate the injected spin current density $j_s$ due to spin pumping. We checked the reproducibility of the voltage measurements by turning the sample (rotation axis is in-plane) by 180° with respect to the dc magnetic field $H_{dc}$, which is equivalent to reversing the $H_{dc}$ direction (Fig. 3a). The normalized voltage $V/h_{rf}^2$ is usually slightly different in the parallel ($H_{dc}$ applied as shown in Fig. 3a) and antiparallel cases, which is due to the sensitive placement of the sample in the electrical field node of the cavity. In such cases, the interfacial charge current density $j_C^{2D}$ were calculated by averaging the voltages obtained in both directions. The measurements were repeated three times at several months intervals, showing similar and reproducible results. The data shown here are the ones recorded on the last run.

**FIGURE CAPTIONS**

**Fig. 1. Description and characterization of NiFe/LAO//STO system**. (a) RHEED oscillations during the growth of the 2 unit-cell LAO film and RHEED diffraction patterns before (left inset) and after (right inset) growth. (b) AFM image of the NiFe/LAO(2 uc)//STO sample surface. (c) STEM cross-section of the NiFe/LAO(2 uc)//STO sample. (d) Temperature dependence of the sheet resistance of a NiFe(2.5 nm)/LAO(2 uc)//STO sample and of a reference NiFe(2.5 nm)//STO sample. The extracted resistance of the 2DES is also plotted, and compared with the measured resistance of a standard LAO(5 uc)//STO sample.

**Fig. 2. Ferromagnetic resonance in NiFe/LAO//STO.** (a) FMR signal at different frequencies at room temperature. (b) Dependence of the FMR resonance frequency on the external magnetic field for a NiFe/LAO//STO sample and a NiFe//LAO reference. (c) Dependence of the spectral peak-to-peak linewidth with the microwave frequency for the same two samples.

**Fig. 3. Spin-to-charge conversion in NiFe/LAO//STO**. (a) Sketch of spin-pumping experimental configuration. Measurements were performed in a cylindrical X-band resonator cavity at 7 K. (b) FMR signal at negative and positive external magnetic fields, at a gate voltage of +125 V. (c) Detected voltage normalized to the square of the amplitude of the rf field (thick solid lines) at negative and positive external magnetic fields at $V_g$=+125 V. The decomposition into symmetric (dotted lines) and antisymmetric (thin solid lines) components is also plotted. Inset: detected voltage for a NiFe//LAO reference. (d) Sketch of a simplified Rashba type system at equilibrium. At the Fermi level there are two Fermi contours with opposite spin textures. (e) Principle of the inverse Edelstein effect: the injection of a current $j_s$ of spins oriented along y ($\sigma_y$) from the ferromagnet creates an accumulation of spin-up ($\sigma_y$ >0) electrons and a depletion of spin-down ($\sigma_y$ <0) electrons ; this shifts the two inequivalent Fermi contours, which generates a transverse charge current along x.

**Fig. 4. Gate control of the inverse Edelstein effect in NiFe/LAO//STO.** (a) Symmetric component of the detected signal for different gate voltages. (b) Gate dependence of the signal amplitude at positive and negative external magnetic fields. (c) Gate dependence of the figure-of-merit of the inverse Edelstein effect $\lambda_{IEE}$. Inset : sketch of the band structure of the 2DES, adapted from [29,31].

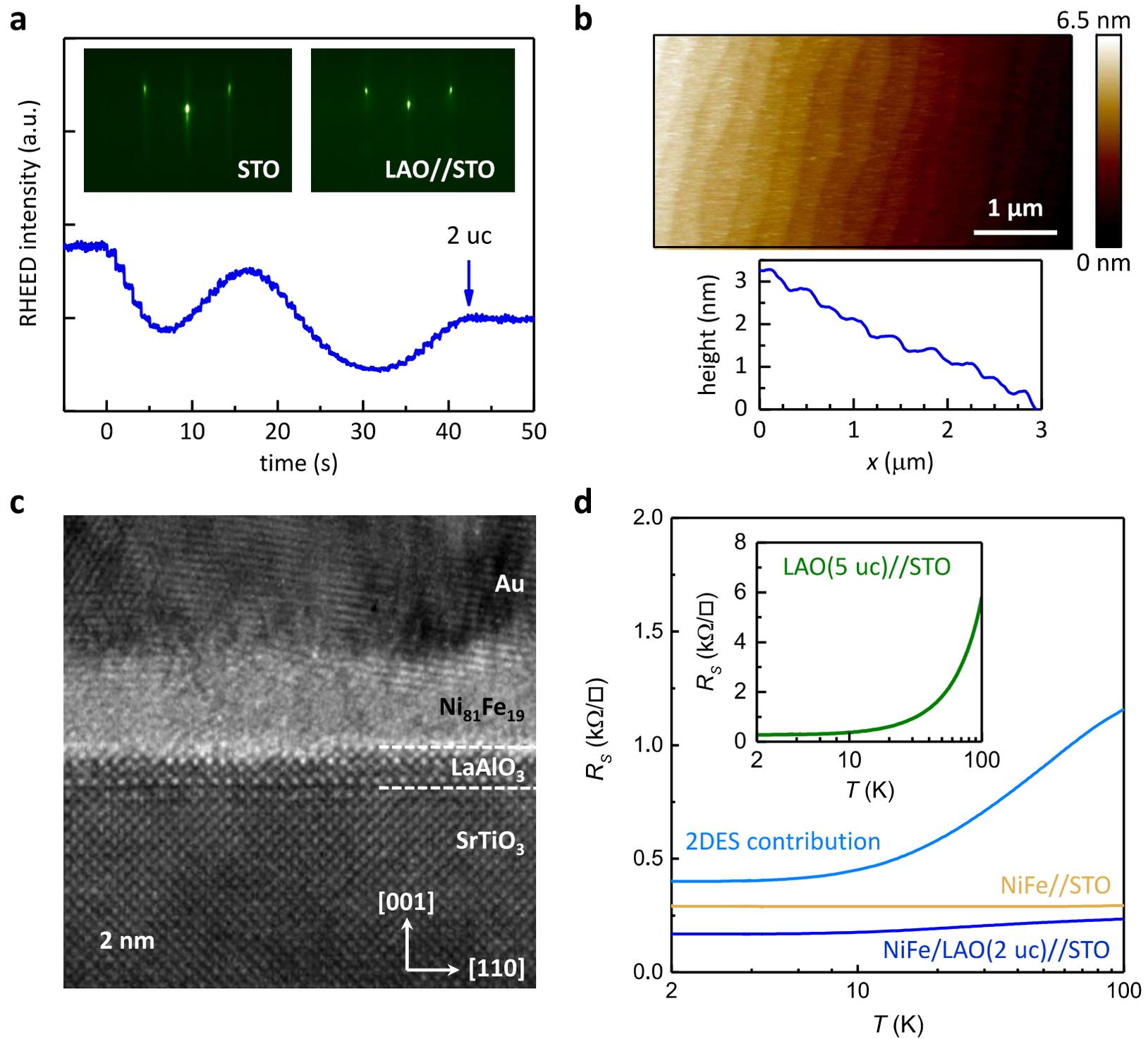

Fig. 1

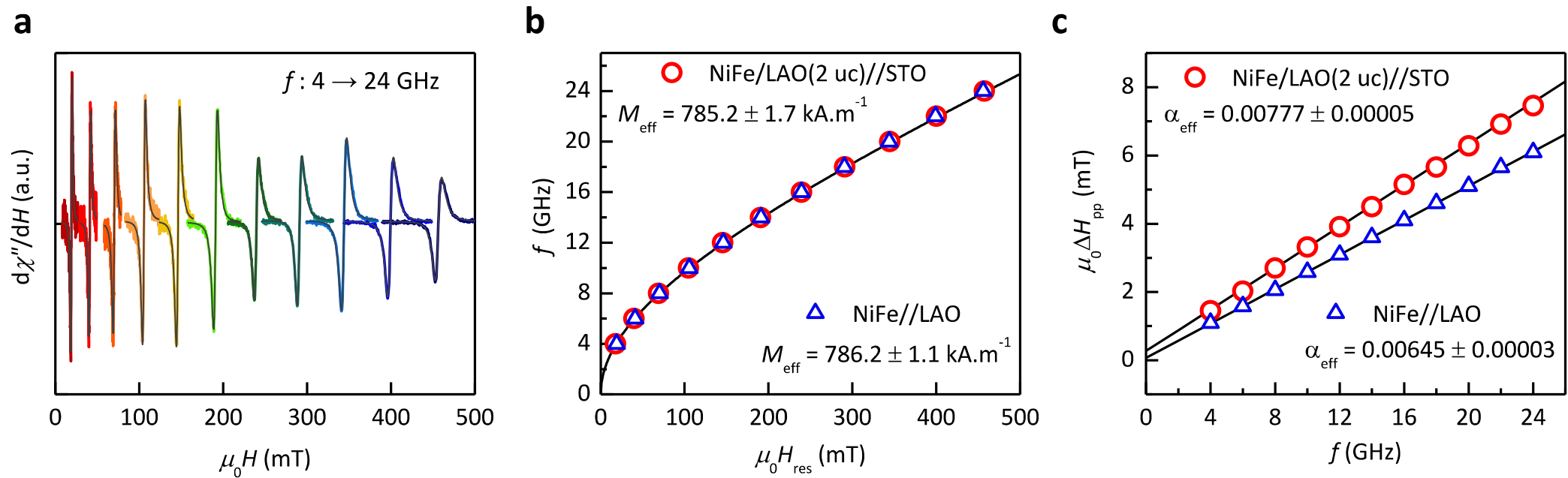

**Fig. 2**

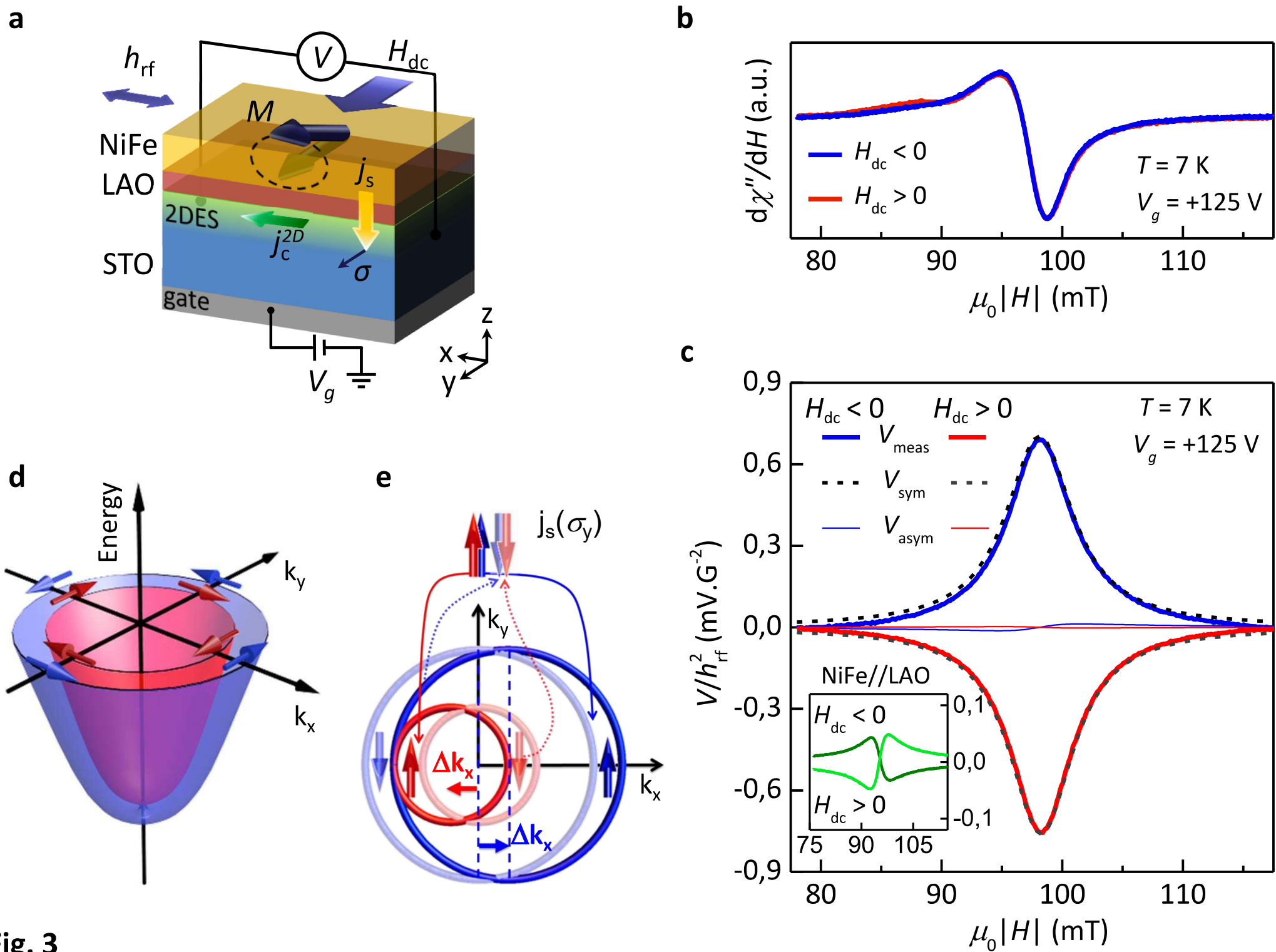

Fig. 3

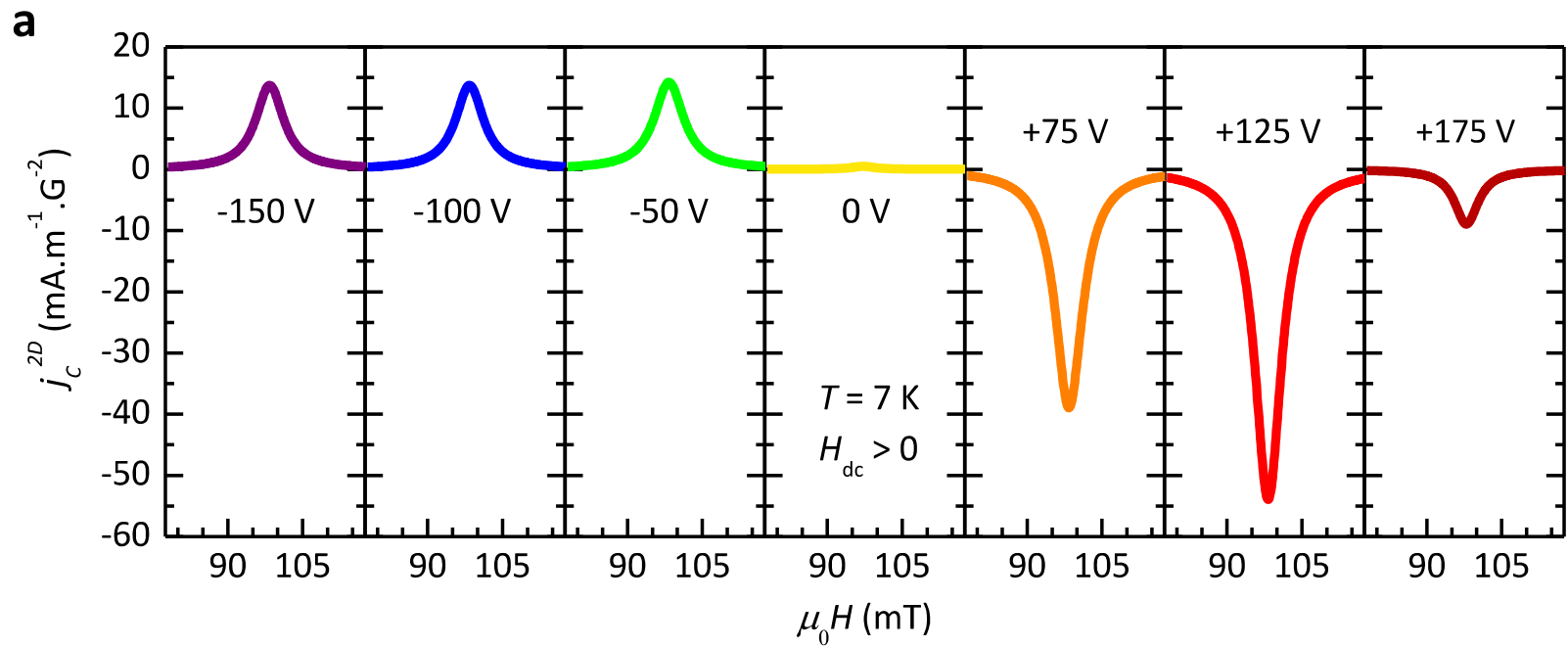
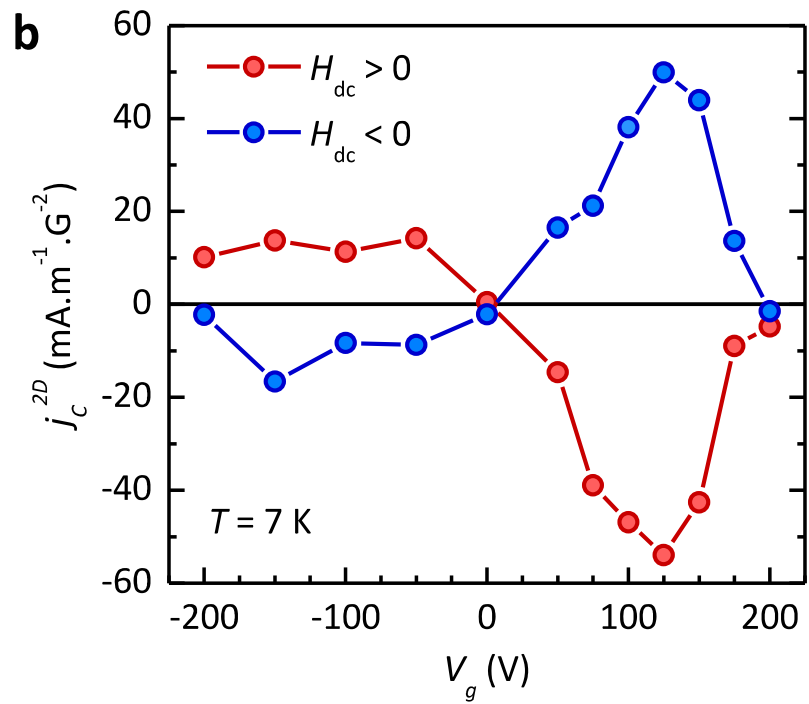
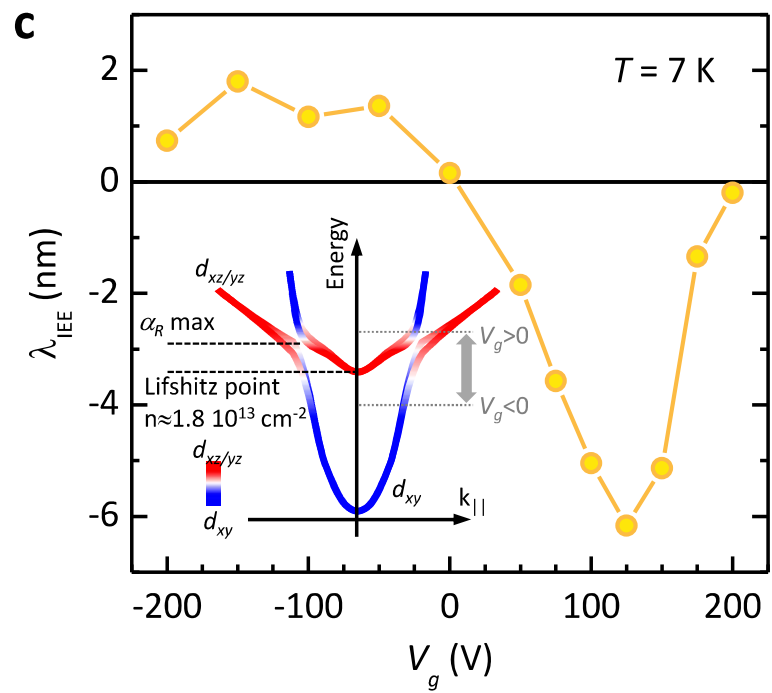

**Fig. 4**